\documentclass[prb]{revtex4}

\usepackage{graphicx}
\usepackage{dcolumn}
\usepackage{bm}
\usepackage{amssymb}

\begin{document}


\title{Nuclear reactor fissile isotopes antineutrino spectra}

\author{V. V. Sinev}
\affiliation{%
Institute for Nuclear Research RAS, Moscow\\
}%


\begin{abstract}
Positron spectrum from inverse beta decay reaction on proton was measured in 1988-1990 as a result of neutrino exploration experiment. The measured spectrum has the largest statistics and lowest energy threshold between other neutrino experiments made that time at nuclear reactors. On base of the positron spectrum the standard antineutrino spectrum for typical reactor fuel composition was restored. In presented analysis the partial spectra forming this standard spectrum were extracted using specific method. They could be used for neutrino experiments data analysis made at any fuel composition of reactor core.
\end{abstract}

\maketitle

\section*{Introduction}

At the end of 80-th there were many experiments in Europe, Russia and the USA 
looking for neutrino oscillations at nuclear reactors [1$-$10]. They were done 
at short distances, from tenths meters up to one km, and did not find oscillations.
Figure 1 shows the ratio of measured antineutrino effect to expected one as a
function of a distance. 

Practically all experiments used inverse beta decay reaction (IBD) on proton to detect reactor antineutrinos: 
\begin{equation}
\bar{\nu_{e}}+p \rightarrow n + e^{+}\\
\end{equation}
IBD reaction products produce characteristic detector response which allows to recognize it between backgrounds. One of the features of reaction (1) is 
that positron energy equals to antineutrino energy minus reaction threshold 
(1.8 MeV). This makes possible to measure antineutrino energy spectrum and obtain some reactor parameters. But to judge on reactor properties one needs to have 
etalon spectrum to compare with the measured one. Previous 
experiments seem could give the etalon spectrum. But not in all [1$-$10] 
experiments the positron spectrum of IBD reaction was measured or it 
was measured with poor statistics. Sometimes, if statistics was enough high it 
is appeared that measurement had high threshold on positrons because of large
background, that rejects soft part of spectrum.

Soft part of positron (antineutrino) spectrum is important for some experiments
looking for oscillations with prevalence of mixing parameter $\theta_{13}$, 
which is under investigation in a number of experiments [11$-$13]. Also this 
part of spectrum contains instabe component caused by the long living isotopes 
like $^{90}$Sr and some other accumulating in the core and containing in the 
spent fuel pool which usually placed in vicinity of reactor.

The experiment having low threshold and high statistics was made at Rovno NPP where at the end of 80-th in former Soviet Union the complex program on neutrino researches was realized under leadership of L. Mikaelyan[6].

In Rovno experiment they measured IBD reaction positron spectrum with threshold of 
150 keV and accumulated 174 000 antineutrino events. Using this statistics firstly
the effect of fuel burn up was experimentally measured [14] through the distortion of antineutrino spectrum caused by the change in fuel composition. At the end of 
70-th in [15] was shown that fissile isotopes spectra should differ and this
makes possible to realize the idea of distant reactor monitoring.

In this paper we present new analysis of Rovno experimental data. As s result
we restored reactor antineutrino spectrum using new algorithm. This analysis 
allowed to separate antineutrino spectrum on four ones belonging to main fissile
isotopes according to antineutrino spectra ratio features given in [16]. 

\begin{figure}
\includegraphics{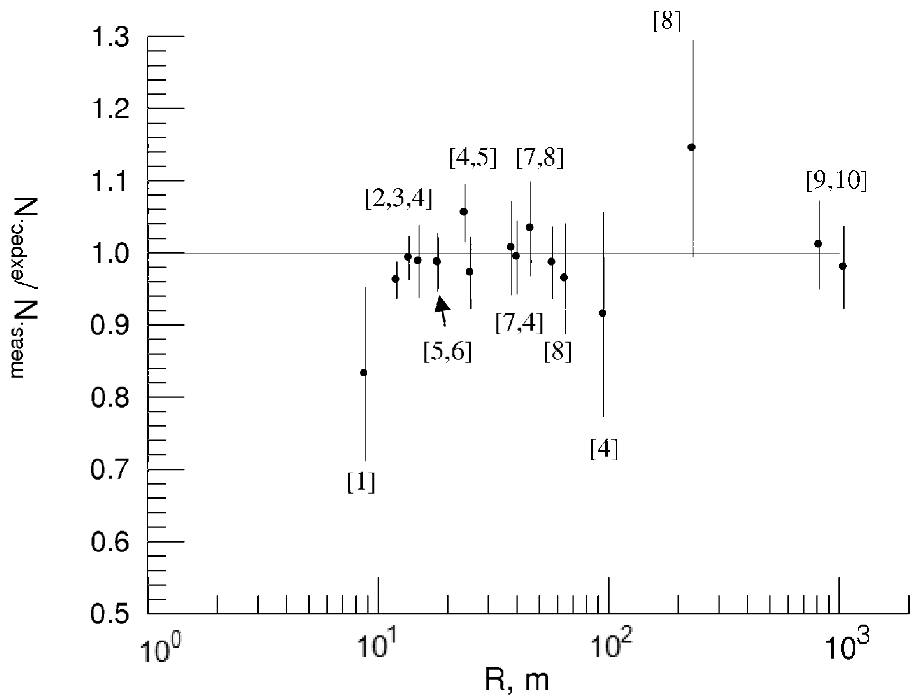}
\caption{\label{fig:fig1} Ratio of measured and expected effects for the number of reactor experiments.}
\end{figure}

\section{Positron spectrum measurement}

They used spectrometer RONS [17], which was constructed having a goal to measure
the positron spectrum originating from the IBD reaction. The special iron 
chamber was builded under the second core of Rovno NPP where the detector was 
installed. Spectrometer contained
1 m$^3$ of liquid scintillator (LS), separated on two detectors: inner (the target, 511 l) and outer (surrounding, 
510 l). The corps containing LS had acrylic walls 2 cm in thickness. Detectors were optically 
isolated and viewed by 84 photomultipliers of PEU-125 type in total.

Experiment was lasted during three nuclear reactor campaigns. One year measurement cycle 
contained of several months measurements before reactor stop on refueling, measurements during 
reactor stop and several months after refueling. When reactor had stopped the 
correlated background was measured. Measured value of correlated background was 215
$\pm$  5 events per 10$^5$ s. 

Antineutrino counting rate was measured with 50\% efficiency. It was 1050 $\pm$ 7 events per 10$^5$ s at full reactor power. This value is mean for all period 
of measurement. As known, the detector counting rate changes during the reactor campaign
on a value about 6\% decreasing from reactor start till its stop on refueling. 
The change of counting rate affected by the change of fuel composition caused
by the burn up effect.

Accidental background was measured in each run and removed on line increasing 
a little bit statistical uncertainty.

Total statistics accumulated during three years is 174 thousand neutrino 
events from IBD reaction on proton.

Long term experiment demands high stability from the spectrometer 
characteristics. Fast spectrum decrease depending on energy and 
slight spectrum deformation (from 3 up to 10\% depending on energy) because of
burn up effect demand exact measurement of detector characteristics.

The number of experiments, which allowed to control with appropriate accuracy the 
measurements, were done during the long term Rovno experiment:
\begin{itemize}
\item Energetic resolution was measured using gamma-sources $^{60}$Co (summed energy 
of two quanta is 2.50 MeV) and $^{24}$Na (summed energy of two quanta is 4.12 MeV). 
Generator of accurate pulses allowed to check differential energy scale non uniformity. 
One time per week smooth spectrum of prompt gammas coming from $^{252}$Cf fissions was measured. 
\item Using gamma sources they checked correctness of accidental background measurement and subtracting it on line. 
\end{itemize}

\section{Antineutrino spectrum}

As a result they measured IBD positron spectrum with high statistics. It is shown on fig. 2. 
This spectrum is presented in visible energy scale, which was taken through detector calibration 
by beta and gamma-sources. Otherwise the spectrum is affected by detector influence so 
called detector response (energetic resolution and energy shift caused by partial absorption 
of annihilation gammas).

\begin{figure}
\includegraphics{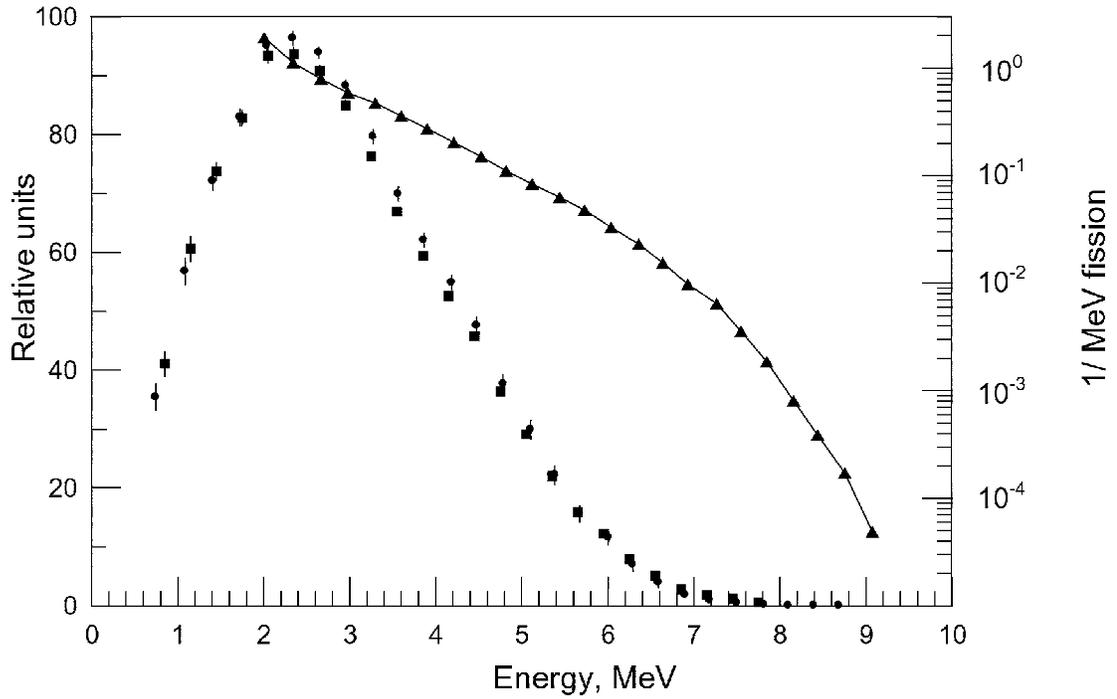}
\caption{\label{fig:fig2} IBD reaction positron spectra. $\blacksquare \ -$ experimental, 
$\bullet \ -$ corrected on the scale and detector response function. 
$\blacktriangle \ -$ experimental restored antineutrino spectrum (right scale).}
\end{figure}

At fig. 3 one can see how ratio of visible energy to real one depends on positron energy [18]. 
This plot was done in KamLAND collaboration on base of analysis of calibration data and 
Monte Carlo simulations. Using this plot we restored real positron energy scale. Also we
 introduced corrections caused by energetic resolution. At fig. 2 the pure positron spectrum 
also shown as well as measured one. Now the purified spectrum can be recalculated in 
antineutrino energy accounting annihilation gammas energy shift. The last procedure applied 
to the spectrum is dividing on IBD cross section. As a result we have reactor antineutrino 
spectrum measured by Rovno experiment at some mean reactor fuel composition which is 
shown on fig. 2 (right scale). Fuel composition corresponding this spectrum is shown in tabl. 1 as well as isotope cross sections.

\begin{figure}
\includegraphics{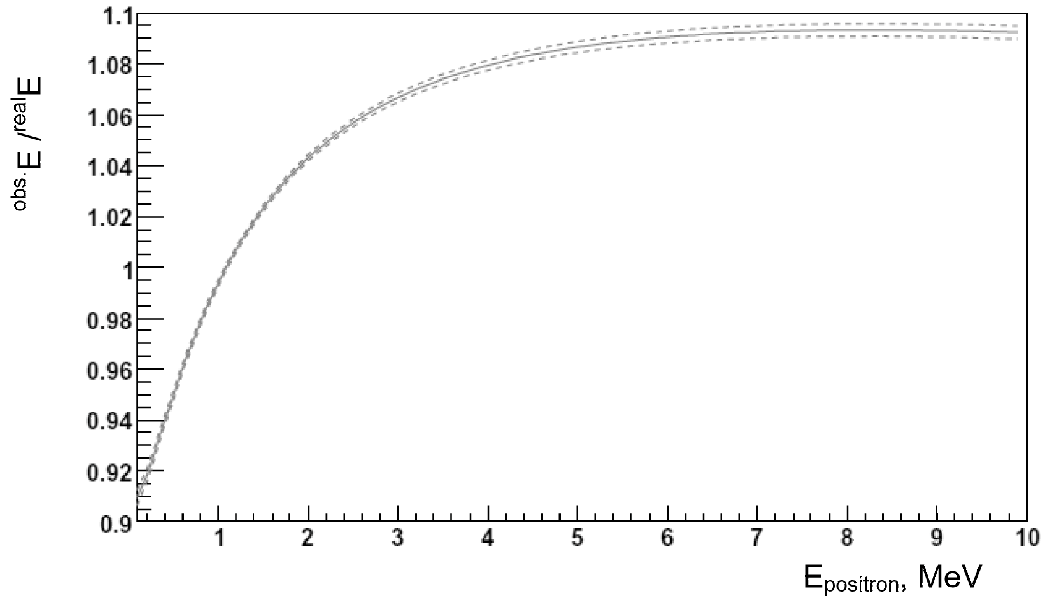}
\caption{\label{fig:fig3} Ratio of visible positron energy to real one. Dashed line marks the error corridor.}
\end{figure}

Received spectrum can be accounted as standard one for nuclear reactors of PWR type 
averaged through the campaign. Remind that during reactor campaign the burn up of uranium 
isotopes ($^{235}$U, $^{238}$U) takes place and in parallel isotopes of plutonium 
($^{239}$Pu, $^{241}$Pu) are
 accumulated in a reactor core. According this the amount of fissions changes because of 
different energy released per fission of different isotopes. Fission energies are shown also
in tabl. 1. 

\begin{table}[h]
\caption{Fuel composition in parts of fission for Rovno spectrum and fission energies for isotopes.}
\label{table:1}
\vspace{10pt}
\begin{tabular}{l|c|c|c|c}
\hline
Isotope & $^{235}$U &  $^{239}$Pu & $^{238}$U &  $^{241}$Pu  \\
\hline
Parts of fiss. & 0.5895 & 0.2893 & 0.0748 & 0.0464 \\
\hline
$E_{fis}$, MeV [25] & 201.92 $\pm$ 0.46 & 209.99 $\pm$ 0.60 & 205.52 $\pm$ 0.96 &  213.60 $\pm$ 0.65  \\
\hline
\end{tabular}\\[2pt]
\end{table} 

\section{Fissile isotopes spectra}

In works [16] by V. Kopeikin was shown that ratio of antineutrino spectra as well as 
beta spectra of fissile isotopes are very stable relative to the method of calculation. It is 
possible to demonstrate that the same is correct for antineutrino spectra received by 
conversion technique [19$-$21]. At fig. 4 one can see spectra ratio for isotopes $^{235}$U and 
$^{239}$Pu from [19$-$25]. One can expect to find the same behavior for ratio of one isotope 
spectrum to the mixture of all. At fig. 5 the ratio of separate spectrum to the mixture from
tabl.1 is shown. Obviously they are stable for spectra taken from a number of 
authors and practically do not depend on the method of how they were taken. 

So, we can propose the procedure of restoring the partial antineutrino spectra basing on these ratios. 
If there is experimentally measured antineutrino spectrum one can apply to it the function of ratio
$k(E)=^{i}S(E)/\sum \alpha_i \ ^{i}S(E)$ if parts of fission are known.

We applied the procedure to the Rovno experimental spectrum and separated it on four fissile 
isotopes spectra. They are shown in tabl. 2. Shown uncertainty accounts experimental statistical 
and systematical uncertainties and uncertainties caused by applied corrections when positron 
spectrum was purified. Fig. 6 shows ratios of fissile isotopes spectra taken from Rovno experiment 
to spectra from [19$-$21] and [24], which are accounted as most accurate for today.  
Cross sections calculated with these spectra appear to be slightly larger than for ILL spectra
 [19$-$21]. Found cross sections are (in brackets the same sections for [19$-$21] spectra):
$^{235}$U $- \ 6.45\times 10^{-43}$ cm$^2$/fiss. (6.40),
$^{239}$Pu $- \ 4.36\times 10^{-43}$ cm$^2$/fiss. (4.19), 
$^{238}$U $- \ 9.14\times 10^{-43}$ cm$^2$/fiss. (8.91),
$^{241}$Pu $- \ 6.28\times 10^{-43}$ cm$^2$/fiss. (5.77). Calculations were made 
for the neutron life time value 885.7 $\pm$ 0.8 s.

\begin{table}[h]
\caption{Fissile isotopes antineutrino spectra (1/MeVfiss.).}
\label{table:1}
\vspace{10pt}
\begin{tabular}{l|c|c|c|c|c}
\hline
$E$, MeV & $^{235}$U &  $^{239}$Pu & $^{238}$U &  $^{241}$Pu & 
$\delta$, \% (68\% C.L.)  \\
\hline
1.75 & 2.957 & 2.537 & 3.444 & 2.991 & 6.8 \\
2.00 & 1.990 & 1.691 & 2.381 & 2.050 & 6.4 \\
2.25 & 1.339 & 1.127 & 1.646 & 1.405 & 5.0 \\
2.5 & 9.762E-01 & 8.097E-01 & 1.231 & 1.035 & 3.2 \\
2.75 & 7.542E-01 & 6.151E-01 & 9.748E-01 & 8.030E-01 & 2.1\\
3.00 & 6.007E-01 & 4.807E-01 & 7.955E-01 & 6.390E-01 & 1.7\\
3.25 & 5.097E-01 & 3.937E-01 & 6.812E-01 & 5.350E-01 & 1.4\\
3.5 & 4.134E-01 & 3.089E-01 & 5.602E-01 & 4.274E-01 & 1.3\\
3.75 & 3.320E-01 & 2.395E-01 & 4.557E-01 & 3.386E-01 & 1.2\\
4.00 & 2.660E-01 & 1.849E-01 & 3.699E-01 & 2.676E-01 & 1.2\\
4.25 & 2.121E-01 & 1.417E-01 & 2.999E-01 & 2.099E-01 & 1.2\\
4.5 & 1.678E-01 & 1.066E-01 & 2.426E-01 & 1.625E-01 & 1.4\\
4.75 & 1.308E-01 & 7.916E-02 & 1.920E-01 & 1.240E-01 & 1.5\\
5.00 & 1.037E-01 & 5.986E-02 & 1.533E-01 & 9.623E-02 & 1.7\\
5.25 & 8.303E-02 & 4.665E-02 & 1.238E-01 & 7.562E-02 & 1.8\\
5.5 & 6.648E-02 & 3.686E-02 & 1.001E-01 & 5.933E-02 & 2.2\\
5.75 & 5.235E-02 & 2.838E-02 & 7.956E-02 & 4.538E-02 & 2.8\\
6.00 & 3.926E-02 & 2.013E-02 & 6.044E-02 & 3.302E-02 & 3.8\\
6.25 & 2.961E-02 & 1.486E-02 & 4.677E-02 & 2.440E-02 & 3.7\\
6.5 & 2.140E-02 & 1.058E-02 & 3.448E-02 & 1.735E-02 & 5.9\\
6.75 & 1.473E-02 & 7.157E-03 & 2.413E-02 & 1.175E-02 & 7.3\\
7.00 & 1.019E-02 & 4.853E-03 & 1.709E-02 & 7.970E-03 & 8.2\\
7.25 & 7.467E-03 & 3.517E-03 & 1.300E-02 & 5.731E-03 & 9.4\\
7.5 & 4.473E-03 & 2.016E-03 & 8.053E-03 & 3.398E-03 & 14\\
7.75 & 2.596E-03 & 1.121E-03 & 4.905E-03 & 2.014E-03 & 20\\
8.00 & 1.350E-03 & 6.328E-04 & 2.792E-03 & 1.136E-03 & 26\\
8.25 & 6.706E-04 & 3.624E-04 & 1.482E-03 & 6.216E-04 & 32\\
8.5 & 3.462E-04 & 2.014E-04 & 7.572E-04 & 3.367E-04 & 40\\
8.75 & 1.837E-04 & 1.011E-04 & 3.994E-04 & 1.771E-04 & 80\\
9.00 & 6.951E-05 & 3.282E-05 & 1.635E-04 & 6.503E-05 & 140 \\
9.25 & 2.930E-05 & 1.267E-05 & 7.213E-05 & 2.599E-05 & 230 \\
\\
\hline
\end{tabular}\\[2pt]
\end{table} 

\begin{figure}
\includegraphics{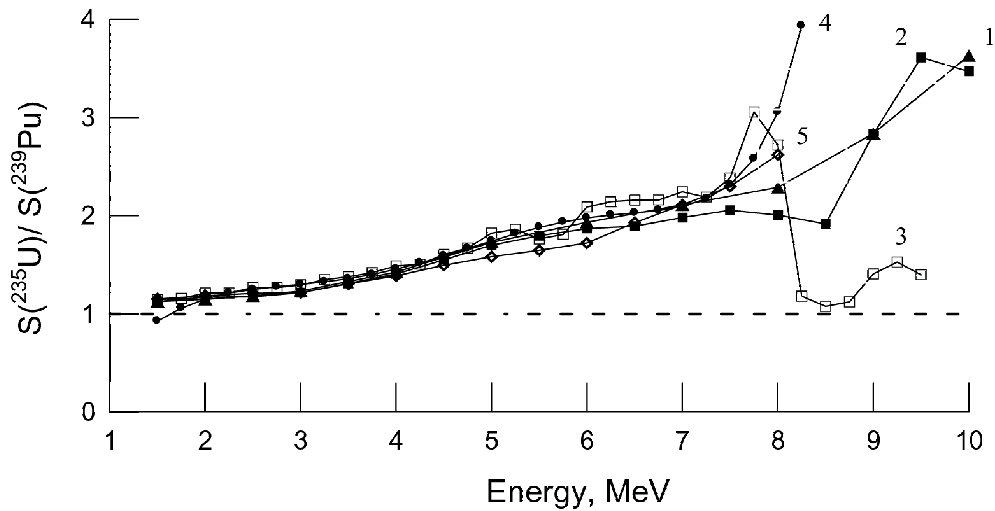}
\caption{\label{fig:fig4} Ratio of $^{235}$U spectrum to $^{239}$Pu: 1 ($\blacktriangle$) $-$ [16], 2 ($\blacksquare$) $-$ [23], 3 ($\square$) $-$ [20, 21],
4 ($\bullet$) $-$ [24], 5 ($\lozenge$) $-$ [22].}
\end{figure}

\begin{figure}
\includegraphics{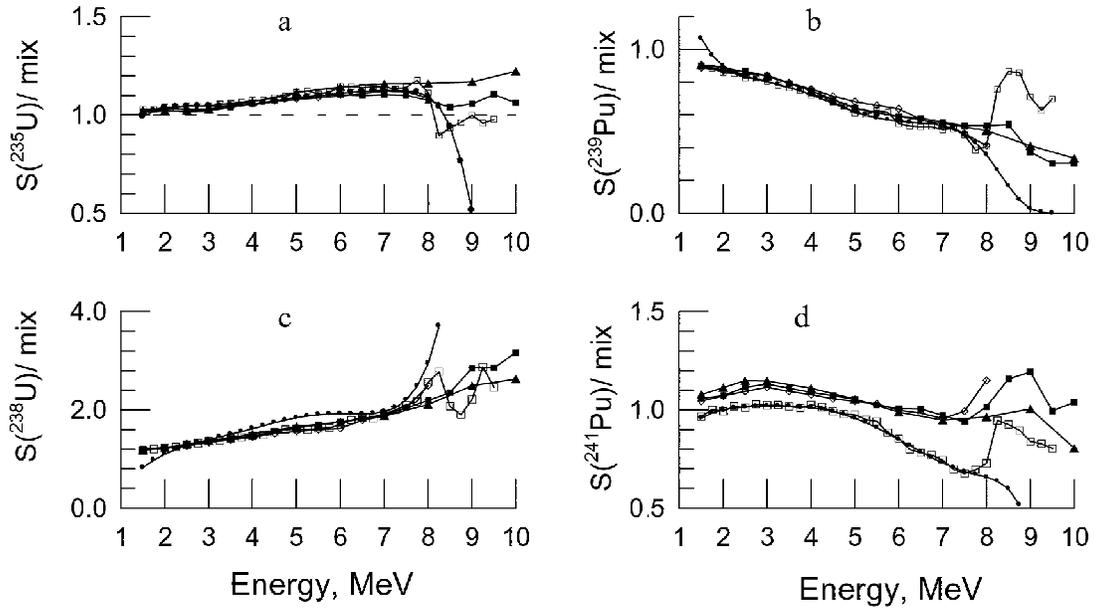}
\caption{\label{fig:fig5} Ratios of fissile isotopes to the mixture from table 1. 
a $- \ ^{235}$U, b $- \ ^{239}$Pu, c $- \ ^{238}$U, d $- \ ^{241}$Pu. 
Marks are the same as on fig. 4.}
\end{figure}

\begin{figure}
\includegraphics{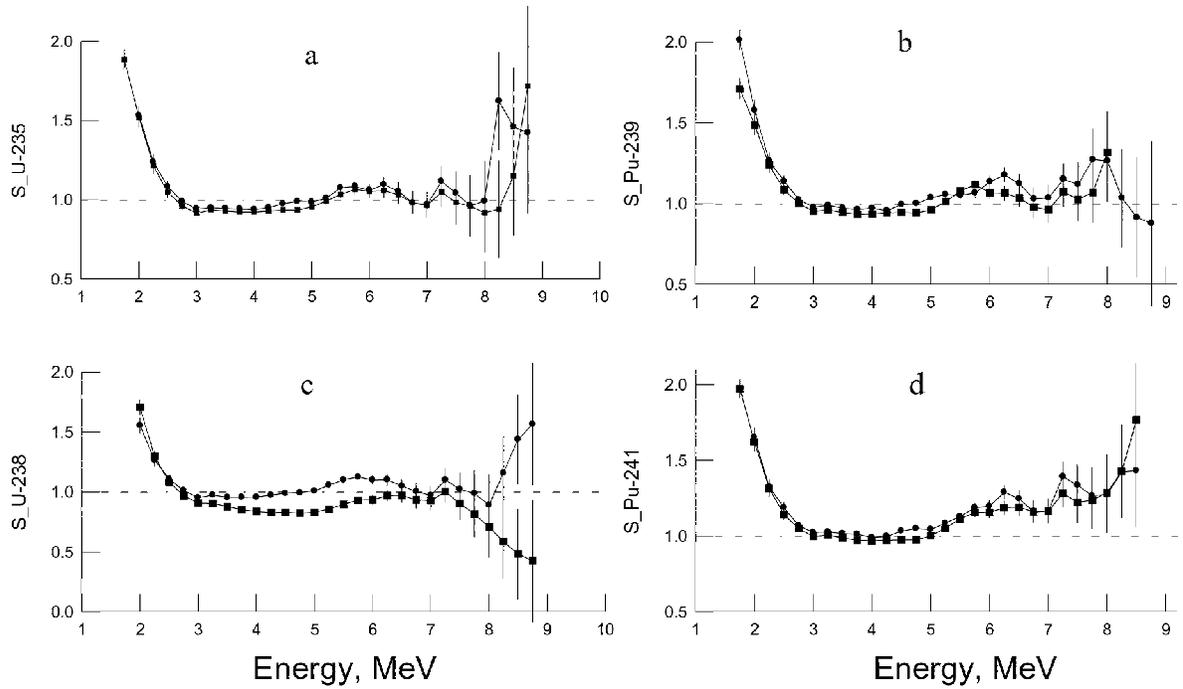}
\caption{\label{fig:fig6} Ratio of Rovno spectra to spectra [19$-$21] ($\bullet$)
 and [24] ($\blacksquare$). a $- \ ^{235}$U, b $- \ ^{239}$Pu, c $- \ ^{238}$U, 
d $- \ ^{241}$Pu.}
\end{figure}

\section*{Conclusion}

At the end of 80-th there was experiment at Rovno NPP measured with high statistics the 
positron spectrum from IBD reaction on proton. In the paper the new analysis is presented. 
As a result new antineutrino spectrum which could be regarded as standard nuclear
reactor spectrum was proposed. 
On base of antineutrino spectra ratios the method of separating experimental spectrum on partial 
ones is proposed. For the first time the partial spectra of fissile isotopes were extracted from 
direct reactor antineutrino measurement. Obtained spectra differ from ILL ones in the soft part were input from long lived isotopes spectra is concentrated. Long lived isotopes are accumulated in the core during the reactor campaign 
and they present also in spent fuel pools close to the reactor.

\section*{Acknowledgment}

Author would like to express acknowledge to L. Mikaelyan for the interest to idea and work and to L. Bezrukov for useful discussions and friendly criticism.

\end{document}